\documentclass[a4paper,11pt]{article}
\usepackage{jheppub}
\usepackage[usenames,dvipsnames]{xcolor}
\hypersetup{
    pdfencoding=unicode,
	colorlinks=true,
	urlcolor=Maroon,
	linkcolor=RoyalBlue,
	citecolor=OrangeRed,
	pdfdisplaydoctitle=true,
	pdfstartview=FitH,
	linktocpage=true
}
\usepackage{tikz}
\usetikzlibrary{shapes.misc}
\newcommand{\yuqi}{\textcolor{purple}}

\title{\boldmath Left-handed string and CHY amplitude at one loop}

\author[a]{Yuqi Li}
\author[a]{and Warren Siegel}
\affiliation[a]{C.~N.~Yang Institute for Theoretical Physics,\\ State University of New York, Stony Brook, NY 11794-3840}

\emailAdd{yuqi.li@stonybrook.edu}
\preprint{YITP-SB-17-46}

\abstract{We propose a generalized left-handed (chiral) gauge choice for the genus one Riemann surface, realized through a singular gauge transformation of worldsheet coordinates. The transformation predominantly affects the logarithmic non-zero modes of the Green's function, leaving non-holomorphic and non-logarithmic modes unchanged. This procedure yields $\delta$-functions for chiral coordinates and box-diagram-like integrals in terms of modular parameters. The resulting $\delta$-functions formulate one-loop level Scattering Equations that simplify to satisfy the tree-level solutions, constraining the locations of the marked points. Subsequent integrals agree with the field-theoretic box diagram for the four-point amplitude, in accordance with the divergent $\epsilon$ expansions derived from dimensional regularization in the infrared limit. We conclude by highlighting potential avenues for future research, including the exploration of methodologies that preclude the need for worldsheet coordinates reparametrization and their implications for accurately capturing infrared behavior from modular parameter integrals.}

\begin{document}
\maketitle
\section{Introduction}
In the last decade, significant efforts have been made to find simpler methods for calculating scattering amplitudes to avoid complicated integrations. The CHY formula, first introduced by Cachazo, He, and Yuan \cite{Cachazo:2013hca,Cachazo:2013iea,Cachazo:2014nsa,Cachazo:2014xea}, is one such method. As a string-like prescription for scattering amplitudes, however, the CHY method is applicable only to tree-level amplitudes of $N$ massless Bosonic particles. It introduces one integral with $N-3$ $\delta$-functions, incorporating so-called Scattering Equations, to determine the locations of each external particle. By solving this series of Scattering Equations, the scattering amplitude is obtained without further integration.

People soon realized that the CHY formula can be derived from certain types of conventional string amplitude with $\delta$-functions inserted \cite{Dolan:2013isa,Bjerrum-Bohr:2014qwa,Dolan:2014ega, Dolan:2015iln}. Mason and Skinner (MS) \cite{Mason:2013sva} introduced the ambitwistor string amplitude with a $\delta$-function inserted in each vertex operator. Berkowitz \cite{Berkovits:2013xba} generalized the pure spinor superstring version of the scattering amplitude by taking the infinite-tension limit ($\alpha ' \rightarrow 0$) while still inserting Dirac $\delta$-functions in the vertex operators. Siegel \cite{Siegel:2015axg} introduced a closed string approach, also known as the left-handed string approach, by applying the singular limit of worldsheet gauge (Hohm-Siegel-Zwiebach (HSZ) gauge \cite{Hohm:2013jaa}) and integrating over the anti-holomorphic variable $\bar{z}$ of standard vertex operators to produce the CHY formula. The supersymmetric generalization was also obtained, and its equivalence to the MS prescription was discussed \cite{Li:2017emw}. Moreover, the recent paper calculates the tensile left-handed string amplitude at tree-level \cite{LipinskiJusinskas:2019cej}.

Since the CHY formula, which is calculated on a Riemann sphere, is only valid at tree-level, the loop-level generalization of the CHY formula has attracted considerable attention. Adamo, Casali, and Skinner constructed the Scattering Equations at one-loop using the Ambitwistor string by transforming a torus into a specially marked Riemann sphere \cite{Adamo:2013tsa}. He and Yuan developed similar Scattering Equations with two massive particles by taking the forward limit \cite{He:2015yua}. The Infrared behavior of one-loop Scattering Equations, primarily focusing on the integrand rather than the final integral, was further explored by Casali and Tourkine \cite{Casali:2014hfa}, in which, however, it is still not so clear how to establish a connection between their work and the Gross-Mende saddle point method \cite{Gross:1987kza, Gross:1987ar}. The link between modular variables and loop momenta for higher loops is considered in a more recent paper \cite{Tourkine:2019ukp}. Further discussions of one-loop integrands on the special Riemann sphere of maximal super-Yang-Mills and supergravity in dimensions other than 10, as well as non-supersymmetric theories, can be found in a more detailed paper \cite{Geyer:2015jch}.

In this paper, following \cite{Siegel:2015axg,Li:2017emw}, we introduce a generalized left-handed (chiral) gauge choice corresponding to the genus one Riemann surface, induced by a singular gauge transformation of worldsheet coordinates. By taking the same singular limit of the geometric gauge parameter ($\beta$) used in tree-level calculations, we generalize the tree-level transformations on worldsheet coordinates to a differential that operates on the corresponding Green's functions \cite{Li:2022tbz}. We find that the predominant effect of the change on the Green's function is on the logarithmic non-zero modes, while the non-holomorphic and non-logarithmic modes, corresponding to the zero modes and a portion of the non-zero modes respectively, remain unchanged. In the singular $\beta\rightarrow\infty$ limit, since the left-handed conditions of the string worldsheet coordinates make the original holomorphic (chiral) and anti-holomorphic (anti-chiral) worldsheet coordinates behave independently, we will reparametrize the worldsheet coordinates such that the anti-chiral part results in both $\delta$-functions for chiral coordinates and the box-diagram-like integrals involving the modular parameters.
Thus, the equations in the $\delta$-functions serve as one-loop level Scattering Equations, the solutions to which are highly degenerate, simplifying to the tree-level solutions in a straightforward manner. As in the tree-level case, these solutions can impose constraints on the locations of marked points. Upon a change of variables, these final integrals corresponding to the gauge multiplets of heterotic string yield the (super-)Yang-Mills result (box diagram for four-point amplitude \cite{Amano:1989vc,Montag:1991hd,DHoker:1994gnm}) of the divergent $\epsilon$ expansions from dimensional regularization in the infrared (IR) limit. Finally, we discuss the potential generalization of our left-handed string approach to avoid using the reparametrization of the worldsheet coordinates, and the corresponding implications for the correct IR behavior resulting directly from the modular parameter integrals.

\section{HSZ gauge at loop-level}
We begin by reviewing the left-handed string approach \cite{Siegel:2015axg} to the tree-level CHY formula. The primary difference between left-handed (chiral) strings and conventional strings lies in the boundary condition of the worldsheet fields. This adjustment effectively changes the sign of the right-handed number operator in the level-matching constraint, thereby eliminating all massive fields. This modification manifests in worldsheet conformal field theory as a change in the sign of the anti-holomorphic part of propagators. For the tree-level case, the conformal gauge has proven beneficial in applying the Kawai-Llewelyn-Tye (KLT) factorization \cite{Kawai:1985xq} with standard open-string amplitudes. The change of boundary condition subsequently leads to an immediate cancellation of $\Gamma$-functions in the four-point amplitude, reproducing the particle result \cite{Huang:2016bdd}. While this process is straightforward for the string amplitude at tree-level, a more comprehensive analysis of zero-mode contributions is necessary for loop-level string amplitudes, as will be given below.

Before we delve into the details, let's first take a step back to review the string amplitude and introduce some relevant notations. Since the left-handed string is essentially a closed string, it is natural to start our discussion with the heterotic string. Consider the vertex operators of the heterotic string \cite{Schlotterer:2016cxa,Mafra:2022wml}, 
\begin{eqnarray}
	V^{a_i}(z_i,\bar z_i)&=&V_{\textrm{SUSY}}(z_i,\epsilon_i,k_i)\bar J^{a_i}(\bar{z}_i)e^{ik_i\cdot X(z_i,\bar z_i)}\nonumber\\
	V^{\text {g }}(z_i, \bar{z}_i) &:=& -\frac{1}{2 \alpha^{\prime}} \tilde\epsilon_i^\mu \bar\partial X_\mu(\bar z_i) V_{\mathrm{SUSY}}(z_i) e^{ik_i \cdot X(z_i, \bar z_i)}
	\end{eqnarray}
with vector indices $\mu=0,1, \ldots, d-1$ in $d$ spacetime dimensions and the insertion point $z_i$ on the worldsheet. The color $a_i$ refers to the $\mathit{SU}(N)$ generator $t^{a_i}$ associated with the $i^{\text {th }}$ external particle with polarization $\epsilon_i$, and momentum $k_i$. The color-dependence of the vertex operators are encoded in the right-moving Kac-Moody currents $\bar J^{a_i}(\bar{z}_i)$, which are replaced by polarization-dependent operators $\tilde\epsilon_i^\mu \bar\partial X_\mu(\bar z_i)$ in the gravity counterpart subject to transversality $(\tilde\epsilon \cdot k)=0$. The left-moving $V_{\mathrm{SUSY}}(z_i)$ corresponds to the vertex operators of the gauge multiplet in the type-I superstring partner without color degrees of freedom. As in \cite{Schlotterer:2016cxa}, we will not distinguish between integrated and unintegrated vertex operators and only consider the symbolic form of the correlation functions of the vertex operators. We refer the reader to \cite{Mafra:2022wml} and \cite{Gerken:2018jrq} for more detailed discussions on the supermultiplets and vertex operators.

\subsection{Green's function}
Following the discussions in \cite{Siegel:2015axg}, we define the original coordinate Hohm-Siegel-Zwiebach (HSZ) transformations:
\begin{eqnarray}\label{eq:def}
\chi:\,z\rightarrow \mathbf{z}=\sqrt{1+\beta}z,\,\, \,\, \bar z \rightarrow \mathbf{\bar z}=\frac{1}{\sqrt{1+\beta}}(\bar z -\beta z).
\end{eqnarray}
After taking the singular limit $\beta\gg 1$, one has:
\begin{eqnarray}\label{eq:limit}
\frac{1}{z}\rightarrow\frac{1}{\sqrt{\beta}}\frac{1}{z}, \,\,\,\,\,
\frac{1}{\bar z}\rightarrow-\frac{1}{\sqrt{\beta}}\frac{1}{z}(1+\frac{1}{\beta}\frac{\bar z}{z}).
\end{eqnarray}
As we will see later, the operator product expansion (OPE) will become chiral in the leading order when taking $\beta\rightarrow\infty$ limit\footnote{The minus sign in $1/\bar z$ limit in (\ref{eq:limit}) will give us one overall $(-1)^N$ factor, with N the number of external legs. We will ignore this factor for the rest of the discussion and absorb this factor in the definition of the tree-level amplitude of generic number of legs. Of course, the sign of the amplitude will not change for our discussion at four-point.}. It is also straightforward to see that, in the HSZ gauge, the Lagrangian changes canonically into the form,
\begin{eqnarray}
\mathcal{L}=-\frac{1}{2}[\beta(\bar\partial X)(\bar\partial X)+(\bar\partial X)(\partial X)].
\end{eqnarray}
Recall that the ordinary Green's function for the genus one Riemann surface is
\begin{eqnarray}\label{eq:Green}
 G(z_{ij}|\tau)=
 -\ln\left|\frac{\theta_1(z_{ij}|\tau)}{\theta'_1(0|\tau)}\right|^{2}+2\pi\frac{(\operatorname{Im}
   z_{ij})^2}{\operatorname{Im}\tau}\,.
\end{eqnarray}
Here, $z_{ij}=z_i-z_j$ and all $\theta_1(z_{ij}|\tau)$s are ordinary Jacobi $\theta$ functions, with $\tau$ the modular parameter of the torus. The $\theta_1(z|\tau)$ function is defined as
\begin{eqnarray*}
\theta_1(z|\tau)=2iq^{\frac{1}{8}}\sin(\pi z)\prod^{\infty}_{j=1}(1-q^j)(1-e^{2\pi i z}q^j) (1-e^{-2\pi i z}q^j)
\end{eqnarray*}
with $q=e^{2\pi i\tau}$ and $\theta'_1(z_{ij}|\tau)$ its derivative. We will further decompose the Green's function by extracting the holomorphic logarithmic terms and keep the leading terms in the expansion of $z$ and $\bar z$:
\begin{eqnarray}
G(z|\tau)&=&-\log|-2\pi i z|^2+2\pi(\frac{\mathrm{Im}[z]^2}{\mathrm{Im}[\tau]}+\mathrm{Im}[z])\nonumber\\
&=&G_{\intercal}(z|\tau)+G_{\bigcirc}(z|\tau)
\end{eqnarray}
with the help of the property of $\theta_1(z|\tau)$ function (as discussed in \cite{Tourkine:2019ukp} as well), 
\begin{eqnarray*}
|\exp\log \theta(z|\tau)|^2&\approx&|\exp(\log(-2\pi i z)+2\pi iz)|^2\nonumber\\
&=&\exp(\log|-2\pi i z|^2-2\pi \mathrm{Im}[z]).
\end{eqnarray*}
And the definition of $G_{\intercal}(z|\tau)$ and $G_{\bigcirc}(z|\tau)$ are\footnote{We change the expressions in terms of $\mathrm{Im}[\bar z_{ij}]$ instead of $\mathrm{Im}[z_{ij}]$ to prepare for the change of variables later. And the symbols denote $\intercal$ for tree-like part and $\bigcirc$ for loop-like part, as we will see their physical meaning later.} 
\begin{eqnarray}
G_{\intercal}(z|\tau)&=&-\log|-2\pi i z|^2\nonumber\\
G_{\bigcirc}(z|\tau)&=&2\pi(\frac{\mathrm{Im}[\bar z]^2}{\mathrm{Im}[\tau]}-\mathrm{Im}[\bar z]).\nonumber
\end{eqnarray}
Note that the HSZ gauge impacts only the analytic logarithmic part of the Green's function, with the remainder left untouched. To understand this, we must revisit another condition in the left-handed string calculation \cite{Siegel:2015axg}, the relative sign change between the holomorphic (chiral) and anti-holomorphic (anti-chiral) parts. For our $G_{\intercal}(z|\tau)$, we apply the relative sign flip to the anti-chiral part, namely,
\begin{eqnarray}\label{eq:flip}
G_{\intercal}(z|\tau)&\overset{\texttt{flip}}{\Longrightarrow}& -[\log(-2\pi i \mathbf{z})\yuqi{-}\log(-2\pi i \mathbf{\bar z})]\nonumber\\
&=&-\log(-2\pi i \sqrt{1+\beta} z)+\log(-2\pi i \frac{1}{\sqrt{1+\beta}}(\bar z -\beta z))\nonumber\\
&\overset{\beta\gg1}{\longrightarrow}&\frac{\bar z}{\beta}\frac{\partial}{\partial z}G_\intercal(z|\tau)=-\frac{\bar z}{\beta z}
\end{eqnarray}
Therefore, the combination of the HSZ gauge and the sign flip leads to a partial differential acting on the analytic part of the Green's function. This modification transforms the Green's function into a $\beta$-dependent form, which we denote as $G_\chi(z|\tau)$, given by
\begin{eqnarray}
G_\chi(z_{ij}|\tau)&=&\frac{\bar z_{ij}}{\beta}\frac{\partial}{\partial z_{i}}G_\intercal(z_{ij}|\tau)+G_\bigcirc(\bar z_{ij}|\tau)\nonumber\\
&=&-\frac{1}{\beta}\frac{\bar z_{ij}}{z_{ij}}+2\pi(\frac{\mathrm{Im}[\bar z_{ij}]^2}{\mathrm{Im}[\tau]}-\mathrm{Im}[\bar z_{ij}]).
\end{eqnarray}
As compared to \cite{Siegel:2015axg}, the transformed logarithmic part (\ref{eq:flip}) appears identical to the tree-level Scattering Equations. Before we demonstrate the equivalence between the two, let us first explore the suitable change of variables that will allow us to isolate the part $G_\bigcirc(\bar z_{ij}|\tau)$, which corresponds to a box-diagram-like integral.

\subsection{Reparametrization of worldsheet coordinates}
As the worldsheet coordinates of the left-handed string are chiral, we will treat the holomorphic (chiral) and anti-holomorphic (anti-chiral) parts separately. We introduce a reparametrization of the worldsheet coordinates as seen in \cite{Amano:1989vc,Montag:1991hd,DHoker:1994gnm}:
\begin{eqnarray}\label{eq:change}
z_i&=&\nu_i\nonumber\\
\bar z_i&=&x_i+\tau y_i,
\end{eqnarray}
with $\tau$ the same modular parameter as discussed in the previous section. In this approach, we treat the anti-chiral variables as real two-dimensional variables, while treating the chiral variables as complex one-dimensional variables. The measure of the conventional string, after setting $\tau_2=\mathrm{Im}[\tau]$ and $\tau_1=\mathrm{Re}[\tau]$, is given by
\begin{eqnarray}\label{eq:measure}
d\mu=d\tau_1 d\tau_2 \prod_{i=1}^3d^2z_i\frac{1}{\tau_2^{d/2}},
\end{eqnarray}
with $d$ the spacetime dimension, which transforms into a form corresponding to this change of variables:
\begin{eqnarray}
d\mu=d\tau_1 d\tau_2 \prod_{i=1}^3dx_i dy_i d\nu_i\frac{1}{\tau_2^{d/2-3}}.
\end{eqnarray}
Here, $x_i$'s and $y_i$'s are all real, and $d\bar z_i=\tau_2dx_idy_i$ is understood. We also further fix $z_4=\tau$ as well as $\bar z_4=\bar\tau$ due to translational invariance on the worldsheet.\footnote{It is worth noting that our approach adopts the same configuration as in \cite{Amano:1989vc,Montag:1991hd}, which differs from the configuration used in \cite{DHoker:1994gnm}.} After the change of variables, we have
\begin{eqnarray}
G_\intercal(z_{ij}|\tau)=-\frac{1}{\beta}\frac{x_{ij}+\tau y_{ij}}{\nu_{ij}}
\end{eqnarray}
for the $G_\intercal(z_{ij}|\tau)$ part, and the $G_\bigcirc(\bar z_{ij}|\tau)$ in the Green's function becomes $G_\bigcirc(y_{ij}|\tau)$ with $y_{ij}=y_i-y_j$, which is
\begin{eqnarray}
G_\bigcirc(y_{ij}|\tau)&=&2\pi(y_{ij}^2-y_{ij})
\end{eqnarray}
The sign in the definition (\ref{eq:change}) of $\bar z_i$ is chosen to make the $G_\bigcirc(y_{ij}|\tau)$ positive definite, which coincides with the definition of the box-diagram-like integral in \cite{Amano:1989vc,Montag:1991hd}.

\section{The amplitude}
The rest of the paper will primarily focus on the (super-)Yang-Mills (SYM) amplitudes that correspond to the vertex operators of the gauge multiplet. The 4-point scattering amplitude at the one-loop level can be represented by
\begin{eqnarray}
A_4&=&\int d\mu\langle V^{a_1}(z_1,\bar z_1)V^{a_2}(z_2,\bar z_2)V^{a_3}(z_3,\bar z_3)V^{a_4}(z_4,\bar z_4)\rangle|_{z_4=\tau}\nonumber\\
&=&\int d\mu\exp(\frac{\alpha'}{2}\sum_{i<j}s_{ij}G(z_{ij}|\tau)) \langle \mathcal{K}(z)\rangle
\end{eqnarray}
where $s_{ij}=k_i\cdot k_j$ are the Mandelstam variables, $d\mu$ is the measure on the string moduli space, and $G(z_{ij}|\tau)$ is the Green's function on a genus-one Riemann surface. The $\langle\mathcal{K}(z)\rangle$ denotes the correlator of the vertex operators with the exponential factors $\exp(ik\cdot X(z))$ stripped off. Given that the short distance behavior arising from the derivative of the Green's function still has simple poles, $\partial z G(z|\tau)\sim\frac{1}{z}$ \cite{Gerken:2020xfv,DHoker:2020prr,DHoker:2023vax,DHoker:2023khh}, the correlator $\langle\mathcal{K}(z)\rangle$ remains chiral in holomorphic ($z$) and anti-holomorphic ($\bar z$) part at the leading order. Therefore, one can factor out the current algebra part, analogous to the tree-level case, and obtain for the single trace factor
\begin{eqnarray}
	J^{a_1}(z_1)J^{a_2}(z_2)J^{a_3}(z_3)J^{a_4}(z_4)\sim\textrm{tr}(t^{a_1}t^{a_2}t^{a_3}t^{a_4})\textrm{PT}(1234)
\end{eqnarray}
with $\textrm{PT}(1234)=\frac{1}{z_{12}z_{23}z_{34}z_{41}}$ representing the Parke-Taylor factor for the ordering $(1,2,3,4)$, as in the tree-level computation. With the help of the KLT relation, the anti-chiral part in a different ordering $(1,3,2,4)$ reads
\begin{eqnarray}
\prod_{i=1}^{4}V_{\mathrm{SUSY}}(\bar z_i,\epsilon_i,k_i)&\sim&\mathrm{S}[1234|1324] A_4^{\mathrm{SYM}}(1,2,3,4)\overline{\textrm{PT}(1324)}\nonumber\\
&=&\frac{i}{\beta^3}\mathrm{S}[1234|1324] A_4^{\mathrm{SYM}}(1,2,3,4)\textrm{PT}(1324)
\end{eqnarray}
Here, we have used (\ref{eq:limit}) in the second line to factor out all the $\beta$ factors\footnote{The application of (\ref{eq:limit}) occurs before fixing any insertion points (i.e. the $z_4$ is fixed in our case). Unlike the tree-level case described in \cite{Li:2017emw}, we extract an additional factor of $i\beta$ caused by fixing of $z_4$ because we will restrict all our integrals for $x_i$ and $y_i$ on the real line later.} in order to make the anti-holomorphic part chiral, and $A_4^{\mathrm{SYM}}(1,2,3,4)$ is the tree-level color-ordered SYM amplitude, with $\mathrm{S}[1234|1324]=\alpha's_{23}$ as the KLT factor \cite{Mafra:2010jq,Broedel:2013tta}. After employing the HSZ gauge and the subsequent change of variables, the amplitude takes the form:
\begin{eqnarray}
A_4&=&\textrm{tr}(t^{a_1}t^{a_2}t^{a_3}t^{a_4}) \mathrm{S}[1234|1324]A_4^{\mathrm{SYM}}(1,2,3,4)\nonumber\\
&\times&\frac{i}{\beta^3}\int d\mu\exp(\frac{\alpha'}{2}\sum_{i<j}s_{ij}G_\chi(z_{ij}|\tau))\mathrm{PT}(1234)\textrm{PT}(1324).\nonumber\\
\end{eqnarray}
Recall the definition of the Green's function and the change of variable above, and then consider the integral as follows:
\begin{eqnarray}
I_4&=&i\frac{\mathrm{S}[1234|1324]}{\beta^3}\int \frac{d\tau_1 d\tau_2}{\tau_2^{d/2-3}} \prod_{i=1}^3dx_i dy_i d\nu_i\exp[\frac{\alpha'}{2}\sum_{i<j}s_{ij}G_\intercal(z_{ij}|\tau)]\\
&\times&\exp[\frac{\alpha'}{2}\sum_{i<j}s_{ij}G_\bigcirc(y_{ij}|\tau)]\mathrm{PT}(1234)\textrm{PT}(1324).
\end{eqnarray}
Since the $x_i$ dependence enters only in the exponential part through $G_\intercal(z_{ij}|\tau)$ but not in the Parke-Taylor factor, we can carry out the integration over $x_i$ first. The $x_i$ integral gives us the following,
\begin{eqnarray}
\int \prod_{i=1}^3dx_i\exp[\frac{\alpha'}{2}\sum_{i<j}s_{ij}G_\intercal(z_{ij}|\tau)]&=&\int \prod_{i=1}^3dx_i\exp[-\frac{\alpha'}{2\beta} (x_{ij}+\tau y_{ij}) \sum_{i<j}\frac{k_i \cdot k_j}{\nu_{ij}}]\nonumber\\
&=&(i\beta)^3\prod_{i=1}^3\delta(\alpha'\sum_{j\neq i}\frac{k_i \cdot k_j}{\nu_{ij}})\exp[-\frac{\alpha'}{\beta} \tau y_i \sum_{i\neq j}\frac{k_i\cdot k_j}{\nu_{ij}}]\nonumber\\
&=&-i(\beta)^3\prod_{i=1}^3\delta(\alpha' \sum_{j\neq i}\frac{k_i \cdot k_j}{\nu_{ij}}).
\end{eqnarray}
In the second line above, we have already used the property of $\delta$-function to eliminate the exponential term proportional to the support of the $\delta$-function and the factor of $2$ is cancelled out by the double summation. The equations in the $\delta$-functions are highly degenerate and each of them are identical to the tree-level Scattering Equations. As in the tree-level CHY formula, we are allowed to fix three of them by introducing a volume of the $\mathrm{SL}(2)$ group. Let's choose the configuration $(\nu_1,\nu_2,\nu_3,\nu_4):=(0,1,z,\infty)$, with $z=-\frac{s_{13}}{s_{12}}$ the solution to the tree-level Scattering Equations in this configuration. The corresponding volume of the $\mathrm{SL}(2)$ group is given by $\mathrm{Vol}(\mathrm{SL}(2))=d\nu_1d\nu_2d\nu_4|\nu_{12}\nu_{24}\nu_{41}|^2$ and the Jacobians corresponding to $\nu_i$ are
$$\mathrm{Jac}(\nu_i)=-\sum_{j\neq i}\frac{s_{ij}}{(\nu^{(0)}_{ij})^2},$$ 
where $\nu^{(0)}_i$ denotes the solution to the Scattering Equations corresponding to the aforementioned configuration. Thus, the integral over $\nu_i$ becomes
\begin{eqnarray*}
&&\mathrm{S}[1234|1324]\mathrm{Jac}(\nu_1)\mathrm{Jac}(\nu_2)\int \frac{\prod_{i=1}^{4}d\nu_i}{\mathrm{Vol}(\mathrm{SL}(2))}\delta(\frac{\alpha's_{13}}{\nu_{13}}+\frac{\alpha's_{23}}{\nu_{23}})\mathrm{PT}(1234)\textrm{PT}(1324)\nonumber\\
&=&\alpha's_{23}(-\alpha'\frac{s_{12}s_{14}}{s_{13}})(-\alpha'\frac{s_{12}s_{13}}{s_{14}})\int \frac{d\nu_3}{\mathrm{Jac}(\nu_3)}\frac{\delta(\nu_3-z)}{\nu_{31}\nu_{32}^2}=(\alpha's_{12})^2\nonumber\\
\end{eqnarray*}
After summing up the contributions from all the permutations, namely the $2\leftrightarrow3$ and $2\leftrightarrow4$ switch for this particular channel, we get $s_{12}^2+s_{14}^2-s_{13}^2=-2s_{12}s_{14}$. Putting everything together, we obtain the following expression for the amplitude:
\begin{eqnarray}
A_4&=&\textrm{tr}(t^{a_1}t^{a_2}t^{a_3}t^{a_4})A_4^{\mathrm{SYM}}(1,2,3,4)\\
&\times&\left[-2\alpha'^2s_{12}s_{14}\right]\int \frac{d\tau_1 d\tau_2}{\tau_2^{d/2-3}}\prod_{i=1}^3dy_i\exp[\pi\alpha'\sum_{i<j}s_{ij}(y_{ij}^2-y_{ij})].
\end{eqnarray}
Note that $\tau_1$ integral is trivial and evaluates to $1$. For simplicity, we further define $A_4^{\text{tree}}=\textrm{tr}(t^{a_1}t^{a_2}t^{a_3}t^{a_4})A_4^{\mathrm{SYM}}(1,2,3,4)$. After making some change of variables
\begin{eqnarray}
u_1&=&y_1,\nonumber\\
u_2&=&y_2-y_1,\nonumber\\
u_3&=&y_3-y_2,\nonumber\\
u_4&=&1-y_3,
\end{eqnarray}
the amplitude simplifies as follows:
\begin{eqnarray}
A_4&=&A_4^{\text{tree}} \left[-2\alpha'^2s_{12}s_{14}\right]\int \frac{d\tau_2}{\tau_2^{d/2-3}}\int_0^1 \prod_{i=1}^4du_i\delta(1-\sum_{i=1}^4u_i)\exp[2\pi\alpha'\tau_2(s_{12}u_1u_3+s_{14}u_2u_4)]\nonumber\\
&=&-2stA_4^{\text{tree}}I_{\text{box}}(s,t)
\end{eqnarray}
Here, we absorbed $\alpha'$ into the definition of $s=-\alpha's_{12}=-\alpha's_{34}$ and $t=-\alpha's_{14}=-\alpha's_{23}$, and the box integral is given by \cite{DHoker:1994gnm,Montag:1991hd}
\begin{eqnarray}\label{eq:box}
I_{\text{box}}(s,t)&=&\int \frac{d\tau_2}{\tau_2^{d/2-3}}\int_0^1 \prod_{i=1}^4du_i\delta(1-\sum_{i=1}^4u_i)\exp[-2\pi\tau_2(su_1u_3+tu_2u_4)]\nonumber\\
&=&-\frac{i\pi^{d/2}\Gamma(-\epsilon)^2\Gamma(\epsilon)}{st\Gamma(-2\epsilon)}[(-s)^{-\epsilon}\,_2F_1(1,-\epsilon;1-\epsilon;1+\frac{s}{t})+(-t)^{-\epsilon}\,_2F_1(1,-\epsilon;1-\epsilon;1+\frac{t}{s})].\nonumber\\
\end{eqnarray} 
Here, $d=4-2\epsilon$ \cite{Smirnov:2006ry}, where $\epsilon$ is the dimensional regulator. As demonstrated in \cite{Li:2022tbz}, the left-handed string, in contrast to traditional formulations of string theory, is not subject to the constraints of critical dimensions. Thus, it is appropriate to define the box integral in $4-2\epsilon$ dimensions. 
\section{Conclusions and Discussions}
In this paper, we have generalized the left-handed (chiral) gauge transformation to the genus one Riemann surface through a singular gauge transformation of worldsheet coordinates. This transformation affects only the logarithmic, non-zero modes of the Green's function and leaves the non-holomorphic and non-logarithmic modes unchanged.  By redefining the anti-chiral variables as
$\bar z_i=x_i+\tau y_i$, the integrals over $x_i$ in this procedure yield the $\delta$-functions for the chiral coordinates. These $\delta$-functions impose constraints analogous to those of the tree-level Scattering Equations. Meanwhile, the integrals over $y_i$ and the modular parameters lead to box-diagram-like integrals.
The relaxation from the constraints of critical dimensions permits us to set our integrals in $4-2\epsilon$ dimensions. Up to some overall factors of $2\pi$, our four-point calculations reproduce the field theory result and are in complete agreement with those of Bern-Dixon-Smirnov (BDS) \cite{Bern:2005iz}.

The final integrals in our calculations are essentially the well-known proper-time integrals, which remain multi-dimensional. Nevertheless, the success of our calculations motivates us to consider whether it is possible to further minimize the number of integrals. In the subsequent work by one of the authoer, we aim to investigate the feasibility of applying our left-handed string approach directly to the modular parameter integrals, eliminating the need for reparametrization of the worldsheet coordinates. Such an approach would align more closely with the original CHY formula, but applicable to Riemann surfaces of arbitrary genus.

\acknowledgments
Y.L. extends heartfelt gratitude to Martin Roček and George Sterman for their generous help. Additionally, Y.L. appreciates the constructive conversations with Yao Ma, Xiaojun Yao, and Peng Zhao. The research of Y.L. and W.S. received the support from the National Science Foundation under Grant No. PHY-2215093.

\bibliographystyle{JHEP}
\bibliography{ref}
\end{document}